%% file: main.tex
\documentclass[12pt,a4paper]{article}
\pdfoutput=1

\usepackage{ifthen} 
\newboolean{pdflatex}
\setboolean{pdflatex}{true} 

\newboolean{articletitles}
\setboolean{articletitles}{true} 

\newboolean{uprightparticles}
\setboolean{uprightparticles}{false} 

\input{preamble}
\usepackage{longtable} 
\usepackage{multirow}
\usepackage{hhline}
\usepackage{lscape}
\usepackage{enumitem}
\usepackage{epstopdf}
\usepackage{gensymb}

\newcommand{\mydecay}{\BsToJPsiPhiPhi}
\newcommand{\myrefdecay}{\BsToJPsiPhi}
\newcommand{\mydec}{\BsToJPsiPhiPhi}

\renewcommand{\thefootnote}{\arabic{footnote}}

\begin{document}

\renewcommand{\thefootnote}{\fnsymbol{footnote}}
\setcounter{footnote}{1}

\input{title-LHCb-PAPER}


\renewcommand{\thefootnote}{\arabic{footnote}}
\setcounter{footnote}{0}



\pagestyle{plain} 
\setcounter{page}{1}
\pagenumbering{arabic}


\section{Introduction}
\label{sec:Introduction}

The study of the \BsToJPsiPhiPhi decay, previously unobserved, allows a precise measurement of the \Bs meson mass and a search for possible resonances in the $\phi \, \phi$ and $\jpsi \phi$ invariant mass spectra, similar to what has been reported for the $B^+ \rightarrow \jpsi\,\phi\,K^+\;$ decay mode~\cite{Lees:2014lra,LHCb-PAPER-2011-033,CDF_Y4140} (the inclusion of charge conjugate
processes is implied throughout).
The most recent theoretical predictions for heavy hadron masses, based on lattice QCD calculations, can be found in Refs.~\cite{Gregory:2010gm, McNeile:2012qf, Lewis:2008fu}.
The current experimental knowledge of the \Bs mass, as summarized in Ref.~\cite{PDG2014}, is dominated by results from the LHCb experiment~\cite{LHCb-PAPER-2011-035}, which were obtained with the \myrefdecay decay using a small fraction of the integrated luminosity collected in the 2010--2012 LHC run.
The \Bs mass measurement using this decay is limited by the precision of the momentum scale.
The \BsToJPsiPhiPhi decay mode is a good alternative to \BsToJPsiPhi since the kinetic energy available to the final-state particles ($Q$-value) is much lower, leading to a 65\% reduction in the systematic uncertainty arising from the precision of the momentum scale.  
\par
The \BsToJPsiPhiPhi decay is also of interest in searches for intermediate states in the \Bs decay chain.
In recent years, many new charmonium or charmonium-like states have been discovered, which are not easily accommodated in the quark model of hadrons~\cite{Swanson:2006st,Klempt:2007cp}.
In a study of $B^+\rightarrow \jpsi\phi \, K^+$ decays, the CDF collaboration reported evidence for a state, in the $\jpsi\,\phi$ invariant mass spectrum, called $Y(4140)$ with mass and width values of $m = 4143.0 \pm 2.9\,(\mathrm{stat}) \pm 1.2\,(\mathrm{syst})~\!\mevcc$
and $\Gamma = 11.7^{+8.3}_{-5.0}\,(\mathrm{stat}) \pm 3.7\,(\mathrm{syst})~\!\mev$~\cite{CDF_Y4140}.
The Belle and BaBar collaborations searched for the $Y(4140)$ using the same \Bp decay mode~\cite{Shen:2009vs,Lees:2014lra} and found no significant signal, although the upper limits on the production rate did not contradict the CDF measurement.
Recently, the D0 collaboration reported a similar structure~\cite{Abazov:2013xda}.
At the LHC, both the LHCb and CMS collaborations have searched for the state in question.
The LHCb collaboration found no evidence with 0.37\invfb of $pp$ collision data~\cite{LHCb-PAPER-2011-033}, in $2.4\,\sigma$ disagreement with the CDF measurement.
A CMS search for the same signature~\cite{Chatrchyan:2013dma} supports the CDF observation.
With two out of five experiments failing to observe the $Y(4140)$ resonance the question of its existence still remains open.
The search for resonances in the  $\phi \, \phi$ invariant mass spectrum is also of interest.
Several experiments have reported a near-threshold enhancement in the  $\phi \, \phi$ invariant mass distribution from the $J/\psi \to \gamma \, \phi \, \phi$ decay~\cite{Ablikim2008ac,PhysRevLett.65.1309,Bisello1986294}.
A partial-wave analysis showed that the structure is dominated by a $0^{-+}$ state called $\eta(2225)$.
This resonance is still controversial and its observation in a different decay mode would be conclusive.
\par 
Theoretical predictions of the \mydecay branching fraction are difficult due to the presence of three vector mesons in the final states.
The $\mydec$ decay is the \Bs counterpart of the measured $B^+ \ra J/\psi\,\phi\,K^+$ and $B^0 \ra J/\psi\,\phi\,K^0$ decays~\cite{Aubert1}.
All these channels are strongly suppressed with respect to the similar decays without the additional $\phi$ meson in the final state.
The suppression factors of the last two channels are $0.048 \pm 0.004$ and $0.057 \pm 0.012$ for the charged and neutral decays~\cite{PDG2014}.
A qualitative comparison with these branching fractions can be done considering that the phase space of the decay \mydecay is smaller by a factor of seven, so the  \mydecay branching fraction is expected to be $\sim 10^{-5}$.
\par
This paper presents the first observation of the decay \mydecay and the decay branching fraction measurement with respect to the reference decay \myrefdecay.
A measurement of the \Bs mass is also presented.
The data sample corresponds to an integrated luminosity of 3.0\invfb  $pp$ collisions collected by the LHCb experiment.
The data were recorded in the years 2011 and 2012 at centre-of-mass energies of 7\tev and 8\tev, respectively.

\section{The LHCb detector}
\label{sec:Detector}

The \lhcb detector~\cite{Alves:2008zz, Aaij:2014jba} is a single-arm forward spectrometer covering the \mbox{pseudorapidity} range $2<\eta <5$, designed for the study of particles containing \bquark or \cquark quarks.
The detector includes a high-precision tracking system consisting of a silicon-strip vertex detector surrounding the $pp$ interaction 
region, a large-area silicon-strip detector located upstream of a dipole magnet with a bending power of about $4{\rm\,Tm}$, and three stations of silicon-strip detectors and straw drift tubes placed downstream of the magnet.
The polarity of the dipole magnet is reversed periodically throughout data-taking.
The tracking system provides a measurement of momentum, \ptot, of charged particles with a relative uncertainty that varies from 0.5\% at low momentum to 1.0\% at 200\gevc.
The minimum distance of a track to a primary vertex (PV), the impact parameter, is measured with a resolution of $(15+29/\pt)\mum$, where \pt is the component of the momentum transverse to the beam, in\,\gevc.
Different types of charged hadrons are distinguished using information
from two ring-imaging Cherenkov detectors. 
Photons, electrons and hadrons are identified by a calorimeter system consisting of scintillating-pad and preshower detectors, an electromagnetic calorimeter and a hadronic calorimeter.
Muons are identified by a system composed of alternating layers of iron and multiwire proportional chambers.
The online event selection is performed by a trigger, which consists of a hardware stage, based on information from the calorimeter and muon systems, followed by a software stage, which applies a full event reconstruction.
\par 
Simulated events are used to determine trigger, reconstruction and selection efficiencies and reconstructed mass distributions.
In addition, simulated samples are used to estimate possible peaking backgrounds from $B$ meson decays that can mimic the $\Bs \to J/\psi \,\phi \, (\phi)$ final states.
In the simulation, $pp$ collisions are generated using \pythia~6~\cite{Sjostrand:2006za} with a specific \lhcb
configuration~\cite{LHCb-PROC-2010-056}.
Decays of hadronic particles are described by \evtgen~\cite{Lange:2001uf}, in which final-state radiation is generated using \photos~\cite{Golonka:2005pn}.
The interaction of the generated particles with the detector, and its response, are implemented using the \geant toolkit~\cite{Allison:2006ve, *Agostinelli:2002hh} as described in Ref.~\cite{LHCb-PROC-2011-006}.

\section{Event selection}
\label{sec:selection}

The final states of the signal and reference channels differ only by the presence of an extra $\phi$ meson in the former mode. 
The selections of the \mydecay and \myrefdecay candidates are done in almost the same way, allowing a partial cancellation of systematic uncertainties in the evaluation of the efficiency ratio.
The \jpsi meson is reconstructed in the $J/\psi \ra \mu^+ \mu^-$ decay while the $\phi$ meson is reconstructed in the $\phi \ra K^+ K^-$ decay.
\par 
Events are selected by the hardware triggers requiring a single muon with transverse momentum \pt $> 1.48\gevc$ or a muon pair with product of transverse momenta greater than $(1.3\gevc)^2$. 
At the first stage of the software trigger, events are selected that contain two muon tracks with \pt $>$ 0.5 GeV/c and invariant mass $m(\mu^+ \mu^-)>2.7\gevcc$, or a single muon track with $\pt > 1 \gevc$ and $\chisqip > 16$ with respect to any PV. 
The quantity \chisqip is the difference between the $\chi^2$ values of a given PV reconstructed with and without the track considered.
The second stage of the software trigger selects a muon pair with an invariant mass that is consistent with the known \jpsi mass~\cite{PDG2014}. 
The decay length significance of the reconstructed \jpsi
candidate, $S_L$, is required to be greater than 3, where $S_L$ is the distance between the \jpsi vertex and the PV, divided by its uncertainty.
\par 
The offline analysis uses a cut-based preselection, followed by a multivariate analysis.
In the preselection all the tracks are required to have a good-quality track fit.
In the $\phi \to K^+K^-$ decay reconstruction, kaons are selected with $p>3\gevc$ and $\pt>200\mevc$, and the vertex is required to have a good-quality fit.
Particle identification (PID) is performed using information from all the subdetectors.
A loose requirement is applied to the PID discriminant of kaons with respect to the pion misidentification $\dllkpi>0$, where $\dllxpi = \ln\mathcal{L}_x-\ln\mathcal{L}_\pi$ is the delta-log-likelihood for the $x$ particle hypothesis with respect to the pion.
For the $\jpsi \to \mu^+ \mu^-$ decay, the two muons are required to have $p>5\gevc$ and to satisfy a loose PID selection, $\dllmupi>-1$.
The invariant mass of the \jpsi candidate is required to be in the interval [3036, 3156]\mevcc, corresponding to a $\pm 4\sigma$ interval around the nominal mass of the \jpsi meson~\cite{PDG2014}.
To select the final $B^0_s \to \jpsi \phi \, (\phi)$ decay, the $\phi$ and \jpsi meson candidates are required to pass the selection cuts $\pt(\phi)>300\mevc$ and $\pt(\jpsi)>400\mevc$, and to form a good-quality displaced vertex.
The collinearity angle, defined as the angle between the reconstructed \Bs momentum and the flight direction determined from the secondary vertex, is required to be smaller than 1.8\degree.
In the \myrefdecay decay selection, to reduce the contamination from non-resonant \BsToJPsiKK decays, the dikaon invariant mass is required to be in the range [980, 1080]\mevcc. 
To improve the mass and decay-time resolutions, a kinematic fit~\cite{Hulsbergen} is applied to both \Bs decays, constraining the mass of the \jpsi candidate to its known value~\cite{PDG2014} and the \Bs momentum to point to the PV.
Finally, the \Bs candidate invariant mass is required to be in the interval [5250, 5490]\mevcc.
\par
Different multivariate selection algorithms, based on a boosted decision tree (BDT)~\cite{Breiman,Roe} with the AdaBoost algorithm\cite{AdaBoost}, are used to select the signal and the reference channel samples. 
The BDT is trained with simulated \Bs samples for the signals, while for the background, a sample of 40 million simulated events containing inclusive $B \to \jpsi X$ decays is used.
For the \mydecay decay channel, the simulated sample is generated according to phase space. 
The BDT input variables are the \pt of the $\phi$ and \jpsi mesons and the vertex $\chi^2$, flight distance significance, $S_L$, collinearity angle and the impact parameter of the \Bs meson with respect to the PV. 
The BDT discriminant threshold is chosen to maximise the figure of merit, 
$\epsilon/(3/2+\sqrt{b})$~\cite{Punzi:2003bu}, where $\epsilon$ is the signal efficiency determined using simulated events and $b$ is the number of expected background candidates estimated using mass sideband events in the data. 
For the \myrefdecay channel the BDT discriminant is selected to maximize $s/\sqrt{s+b}$, where $s$ and $b$ are the expected signal and background yields, estimated from simulated events and
sideband data, respectively.
\par
In the \mydecay selection, no restriction is initially put on the $K^+ K^-$ system invariant mass, with both the resonant and the non-resonant $B^0_s \ra \jpsi\, (K^+ K^-)\,(K^+ K^-)$ selected.
If the candidate $B^0_s \ra \jpsi (K_1^+ K_1^-)\,(K_2^+ K_2^-)$ passes the selection cuts, almost always the candidate $B^0_s \ra \jpsi (K_1^+ K_2^-)\,(K_2^+ K_1^-)$ also passes the cuts, resulting in a duplicated candidate.
So a genuine resonant $B^0_s \ra \jpsi \phi(K^+ K^-)\,\phi(K^+K^-)$ event will most of the time produce also a ``fake'' non-resonant candidate, given the low probability that the invariant mass of two wrongly-coupled kaons is around the $\phi$ mass.
In order to remove these ``fake'' candidates, the $K^+ K^-$ system masses are required to satisfy $| m(K^+K^-) - m_\phi | < 15\mevcc$.
After this cut, 1.8\% of events contain double candidates.
For each of these events, one candidate is chosen at random.
In the \myrefdecay decay selection, this ambiguity problem is not present, so a tight cut on the $| m(K^+K^-) - m_\phi |$ is not applied.

\section{Results}
\label{sec:results}
The \mydecay decay branching fraction is measured with respect to the reference decay \myrefdecay as
\begin{equation*}
  \label{eq:mainequation}
  \frac{\mathcal{B}(\BsToJPsiPhiPhi)}
       {\mathcal{B}(\BsToJPsiPhi)}=
       \frac{N_{\mathrm{obs}}(\BsToJPsiPhiPhi)}
            {\epsilon(\BsToJPsiPhiPhi)}
            \cdot \frac{\epsilon(\BsToJPsiPhi)}
                  {N_{\mathrm{obs}}(\BsToJPsiPhi)}
                  \cdot \frac{1}
                        {\mathcal{B}(\phi \rightarrow K^+K^-)}\,,
\end{equation*}
where $N_{\mathrm{obs}}(\BsToJPsiPhiPhi)$ and 
$ N_{\mathrm{obs}}(\BsToJPsiPhi)$ are the numbers of observed events and  $\epsilon(\BsToJPsiPhiPhi)$ and $\epsilon(\BsToJPsiPhi)$ are the selection efficiencies.
\par
\begin{figure}[!htbp]
  \centering
  \includegraphics[width=0.6\textwidth]{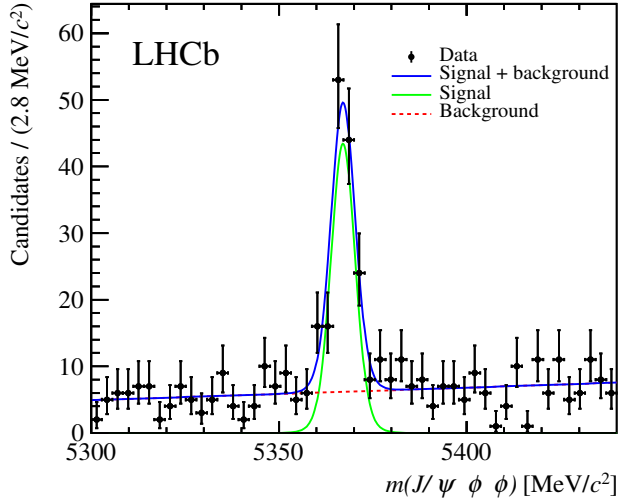}
  \caption{Invariant mass of reconstructed \mydecay candidates. The result of the fit to the distribution is also shown.}
  \label{fig:BsMassJpsiPhiPhidata}
\end{figure}

Figure~\ref{fig:BsMassJpsiPhiPhidata} shows the invariant mass of the reconstructed \mydecay decay, for all the candidates surviving the pre-selection, the BDT and the selection on the $m(K^+K^-)$ around the $\phi$ mass.
In order to evaluate the number of signal decays, an unbinned extended maximum likelihood fit is performed assuming a Gaussian signal peak and an exponential combinatorial background.
The observed signal yield is $128\pm 13$ events, where the uncertainty is statistical.
Using Wilks's theorem~\cite{Wilks:1938dza}, the statistical significance is found to be 15 standard deviations.
As expected, the mass resolution is good, $\sigma = 3.05 \pm 0.41$\mevcc, due to the low $Q$-value of the decay.

\par
\begin{figure}[!b]
  \begin{minipage}[t]{0.5\textwidth}
    \centering
    \includegraphics[width=1.0\textwidth]{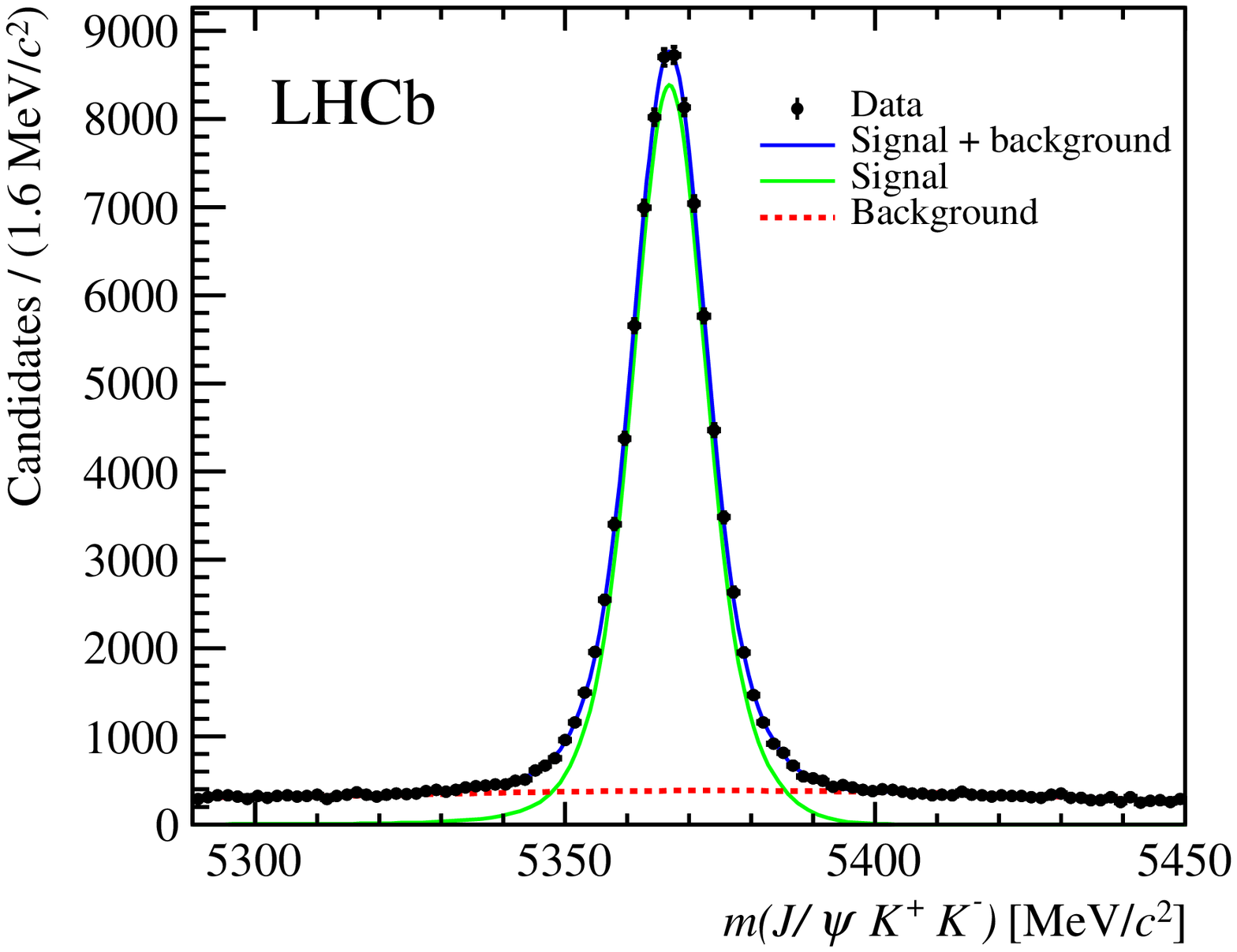}
  \end{minipage}
  \begin{minipage}[t]{0.5\textwidth}
    \centering
    \includegraphics[width=1.0\textwidth]{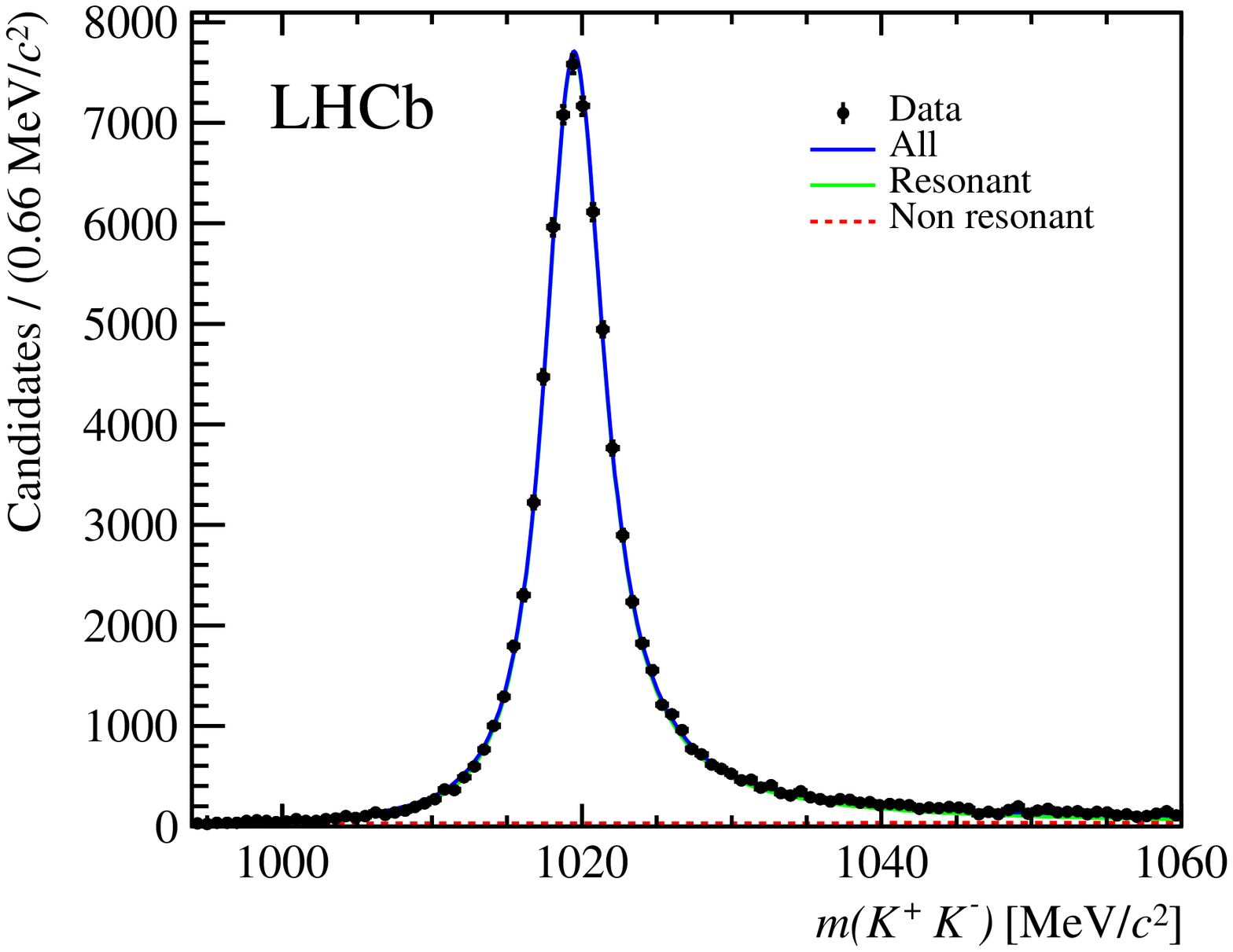}
  \end{minipage}
  \caption{(left) Invariant mass distribution of reconstructed $B^0_s \to J/\psi \, K^+ K^-$ candidates;
    (right) invariant mass distribution of the $K^+ K^-$ system, for those candidates which come from a \Bs decay.
    For both distributions, the fit result is shown.}
  \label{fig:MassesJpsiPhidata}
\end{figure}

In the reference channel, in order to discriminate between the resonant \myrefdecay and the non-resonant \BsToJPsiKK decays, a fit to the $K^+K^-$ mass spectrum is made, where the combinatorial background in the \Bs mass window is statistically removed using the \sPlot~technique~\cite{splot}.
Figure~\ref{fig:MassesJpsiPhidata} (left) shows the mass distribution of the \BsToJPsiKK candidates with the fit results superimposed. 
The \Bs peak is described by a double Crystal Ball function~\cite{Skwarnicki:1986xj}, while the underlying combinatorial background is described by an exponential function plus a second-order polynomial.
Figure~\ref{fig:MassesJpsiPhidata} (right) shows the $K^+K^-$ mass distribution superimposed with the result of the fit.
The fit is performed using a relativistic Breit-Wigner convolved with a Gaussian resolution function for the \Pphi signal and a second-order polynomial for the non-resonant component.
The observed \myrefdecay yield is $82120\pm330$ events where the uncertainty is statistical.

All the efficiencies (detector acceptance, reconstruction, trigger and selection) are evaluated using the simulated samples together with a data-driven method~\cite{TrackingPaper} for tracking and PID.
To check the reliability of the simulation, a comparison is made between data and simulation for all of the kinematic variables used in the selection; good agreement is found.
Since the ratio of $\epsilon$(\mydecay) over $\epsilon$(\myrefdecay) is evaluated, many systematic effects, related to possible small deviation of simulation with respect to data, cancel or are significantly reduced.
\par

The efficiency ratio $\epsilon(\BsToJPsiPhiPhi)/\epsilon(\BsToJPsiPhi)$ is evaluated to be $0.2778\pm0.0015$, where the uncertainty is statistical, due to the limited simulated sample sizes.
As expected, the efficiency of the \mydecay channel is lower than that of \myrefdecay one, due to the presence of the additional $\phi \to K^+ K^-$ decay and the fact that on average the decay products have a smaller transverse momentum.
\par 
From the event yields and the ratio of efficiencies, and using the known $\phi \to K^+K^-$ branching fraction~\cite{PDG2014}, the branching fraction ratio is measured to be

\begin{equation*}
  \frac{\mathcal{B}(\BsToJPsiPhiPhi)}
       {\mathcal{B}(\BsToJPsiPhi)}=
       0.0115 \pm 0.0012\;(\mathrm{stat}) \;^{+0.0005}
_{-0.0009}\;(\mathrm{syst})\,.
\end{equation*}

\noindent
The systematic uncertainty will be discussed in Section~\ref{sec:systematics}.
\par
From the fit to the \Bs invariant mass distribution in the \mydecay decay, the mass of the \Bs meson is measured to be

\begin{equation*}
 m(B^0_s) = 5367.08 \pm 0.38\; \mathrm{(stat)} \pm 0.15 \; \mathrm{(syst)}\mevcc.
\end{equation*}

\begin{figure}[!htbp]
  \centering
  \includegraphics[width=1.0\textwidth]{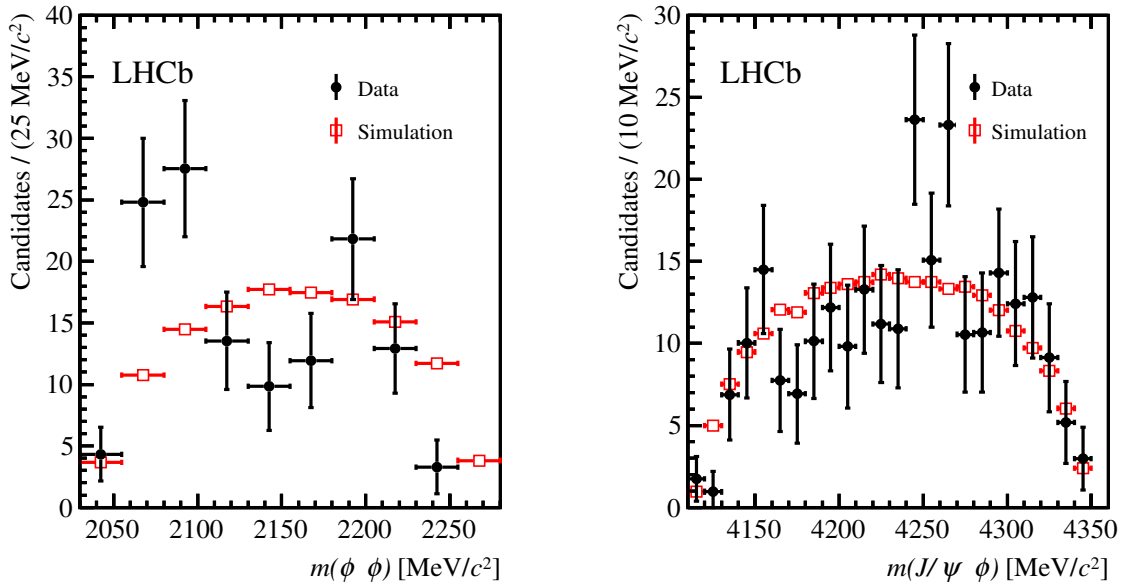}
  \caption{Invariant mass of the ($\phi,\phi$) ({left}) and ($J/\psi,\phi$) ({right}) pairs, in the \mydecay decay.
  In the $J/\psi \,\phi$ plot, for each candidate two values are calculated, one for each $\phi$ meson. 
  The distribution of data and simulation (phase space) are shown.
  To compare the shape, the two distributions are normalised to the same area.}
  \label{pair_masses}
\end{figure}

The $\jpsi \phi$ and $\phi \, \phi$ mass distributions are shown in Fig.~\ref{pair_masses} for both data and simulation.
For the data, the \sPlot~techique is used to subtract the background from the signal.
Since the \mydecay process is a decay of a pseudoscalar into three vector mesons, its accurate description is complex and affected by large theoretical uncertainty.
Here, to simulate the \mydecay decay, a simple phase-space decay model is used, which turns out not to provide a satisfactory description of the data.
The disagreement can be due to either intermediate resonances or the simplified description of the decay.
More data are needed to resolve the issue.
Presently, due to the low statistics and the unknown decay dynamics, it is difficult to draw any conclusions from the two mass distributions.

\section{Systematic uncertainties}
\label{sec:systematics}

A summary of the systematic uncertainties on the measurement of the branching fraction ratio is given in Table~\ref{tab:systsum}.
Since the various effects are uncorrelated, the total systematic uncertainty is evaluated by adding all terms in quadrature.
\par
The average multiplicity of \mydecay candidates in the simulated sample is 1.006 compared to 1.018 of the data.
The relative difference (1.2\%) is assigned as a systematic uncertainty.
\par
The use of a simplified decay model affects the determination of the detection efficiency and introduces some bias in the measurement.
In order to evaluate the effect, the simulated sample is used to study the efficiency of the selection as a function of the two masses $m(\phi,\phi)$ and $m(J/\psi,\phi)$.
The efficiency is then evaluated in a simulated sample reweighted in such a way as to reproduce the mass distributions in the data.
A relative difference $\Delta\epsilon/\epsilon = 1.0\%$ is found and is assigned as a systematic uncertainty due to the unknown decay model.
\par

Alternative functions for describing the signal component are tested: double Gaussian or double Crystal Ball function for the \myrefdecay and single Gaussian or single Crystal Ball function for \BsToJPsiPhiPhi.
In both cases a negligible change in yields is observed and therefore no systematic uncertainty is assigned.
Conversely, different choices of the background parametrisation in the
\myrefdecay data can lead to sizeable difference in the results.
In order to estimate a systematic uncertainty, the fits are repeated using an exponential, a second-order polynomial and the sum of the two (the nominal fit).
The largest difference in yield, 1.6\%, between the nominal fit and the fit with the exponential, is taken as the systematic uncertainty.
The same procedure applied to the signal channel results in a 0.8\% change in yield. The two systematic uncertainties are added in quadrature to give an overall uncertainty of 1.8\% in the modelling of the signal and backgrounds.
\par 

To evaluate the contamination from non-resonant \BsToJPsiPhiKK and \mbox{$B^0_s \to J/\psi \kern0.12em K^+K^- K^+ K^-$} decays, a dedicated search is performed for these two channels in the whole allowed kinematic region (without any requirement on the $K^+K^-$ mass).
The yields are then extrapolated to the restricted kinematic region of the signal.
For the $B^0_s \to J/\psi \, K^+K^- K^+ K^-$ decay, the \sPlot technique is first used to select the \Bs decay and then the two $K^+K^-$ mass spectra are fitted simultaneously to determine the yield of the fully resonant decay candidates and the non-resonant ones.
The non-resonant component is the sum of true non-resonant decays plus the candidates obtained by exchanging the kaons pairings in the resonant decays.
When the latter component is subtracted from the measured yield, the number of non-resonant candidates is found to be $22 \pm 18$.
Extrapolating this number to the $\phi$ meson mass region and using the Feldman-Cousins method~\cite{FeldmanCousins} gives an upper limit of 1.5 events in the signal region at 68.3\% confidence level.
A similar procedure is followed for the $B^0_s \to J/\psi \, \phi \, K^+ K^-$ decay.
One $K^+K^-$ pair is required to have the mass in the non-resonant range, $m(K^+K^-)>1080\mevcc$.
In these events, no evidence of a mass peak is found in the mass spectrum $m(J/\psi \, K^+K^-K^+K^-)$ nor in the mass spectrum of the other kaon pair.
Using the Feldman-Cousins method an estimated contamination of 6.2 events is found at 68.3\% confidence level.  
The uncertainties on the two non-resonant modes are added linearly, resulting in an asymmetric relative uncertainty of $-6\%$.
\par 

The data-driven method used to correct the tracking efficiency for the two additional kaons in the final state of \mydecay with respect to \myrefdecay decay has an uncertainty of 1.5\% per track, resulting in an overall relative uncertainty of 3.0\%.
This term also takes into account the uncertainty of hadronic interactions in the detector material.
\par 

Due to the decay time requirement on the selected events, the lack of knowledge of the admixture of $B^0_{sH}$ and $B^0_{sL}$ eigenstates in the \mydecay decay is a further source of systematic uncertainty~\cite{DeBruyn:2012wj}.
While for the \myrefdecay decay the simulation uses the measured fractions of $B^0_{sH}$ and $B^0_{sL}$ states~\cite{Bs_JpsiPhi}, the \mydecay decay is simulated assuming a completely symmetric combination.
In order to evaluate the systematic effect, the simulated sample is reweighted assuming the two extreme cases where the $B_s$ meson is a complete $B^0_{sH}$ or a $B^0_{sL}$ state.
The observed difference in the efficiency is $2.1\%$ and this number is assigned as the systematic uncertainty. 
\par

A detailed comparison between data and simulation is performed for all the variables used in the BDT selection.
For both the \mydecay and \myrefdecay decay channels, all variables show good agreement and the relative branching fraction result is stable against changes in the threshold of the BDT response.
The total systematic uncertainty on the ratio of branching fractions is found to be $^{+4.4}_{-7.4}\,\%$.

\begin{table}[t]
  \centering
  \caption{Summary of the relative systematic uncertainties (in percentage) on the branching fraction ratio measurement.}
  \begin{tabular}{lc}
    Source & Value \\ \hline
    Candidate multiplicity & $\pm 1.2$ \\
    Decay model & $\pm 1.0$ \\
    Signal and background modelling & $\pm 1.8$ \\
    Contamination from non-resonant decays & $-6.0$ \\
    Track efficiency & $\pm 3.0$ \\
    $B^0_{sH}/B^0_{sL}$ fractions & $\pm 2.1$ \\ \hline
\rule{0pt}{2.5ex}  
    
    	Total & $^{+4.4}_{-7.4}$ \\
  \end{tabular}
  \label{tab:systsum}
\end{table}

Table~\ref{tab:Msystsum} gives a summary of the systematic uncertainties of the \Bs mass measurement.
For the \Bs mass determination, the momentum scale calibration is the main source of systematic uncertainty. 
The momentum scale takes into account the limited knowledge of the detector alignment.
By comparing measured mass values for several charmed mesons with precisely known values, an uncertainty of 0.03\% on the momentum scale is estimated~\cite{TrackingScale}.
The corresponding uncertainty in the \Bs mass value is $\pm0.12\mevcc$.
\par

The uncertainty in the kaon mass~\cite{PDG2014} will affect the \Bs mass determination, while the uncertainty on the \jpsi mass has a negligible effect.
The effect is estimated by repeating the fit with the kaon mass shifted by $\pm \sigma$, where $\sigma$ is the uncertainty on the known kaon mass.
The observed mass variation, $\pm0.06\mevcc$, is assigned as a systematic uncertainty.
\par 

The fit model for the signal and background of the invariant mass distributions is another source of systematic uncertainty.
The effect is estimated by comparing to the nominal case the fit results with those from alternative functions.
The systematic uncertainty from this effect is $\pm 0.02\mevcc$.
\par

The energy loss of the kaons in the detector is another possible source of bias in the mass measurement.
A detailed study of this effect has been performed in Ref.~\cite{LHCb-PAPER-2011-035} for the \myrefdecay decay.
Following the same procedure in \mydecay decay, the effect is found to be $\pm0.06\mevcc$, which is assigned as a systematic uncertainty.
The bias for neglecting the QED radiative corrections in the final state is negligible due to the restricted phase space~\cite{LHCb-PAPER-2011-035}.
The uncertainty due to detector alignment is also negligible.
Combining all of the above sources in quadrature, the total systematic uncertainty on the mass measurement is found to be $\pm 0.15\mevcc$.
\par

As a cross check, the mass measurement is performed separately in the two data-taking periods and in two samples with opposite magnet polarity.
All the measurements are consistent within the uncertainties.

\begin{table}[h]
  \centering
  \caption{Summary of the absolute systematic uncertainties (in \mevcc) affecting the \Bs mass determination from the \mydecay decay.}
  \begin{tabular}{lc}
    Source & Value \\ \hline
    Momentum scale & 0.12 \\
    Kaon mass & 0.06 \\
    Kaon energy loss & 0.06 \\
    Signal and background modelling & 0.02 \\ \hline
    	Total & 0.15 \\
  \end{tabular}
  \label{tab:Msystsum}
\end{table}

\section{Conclusions}
\label{sec:conclusions}
This paper presents the first observation of the $\mydec$ decay channel, with a signal yield of $128\pm13$.
Taking the \myrefdecay decay as the reference channel the relative branching fraction is measured to be

\begin{equation*}
  \frac{\mathcal{B}(B_s\rightarrow J/\psi\,\phi\,\phi)}
       {\mathcal{B}(B_s\rightarrow J/\psi\,\phi)}=
       0.0115\pm 0.0012\;(\mathrm{stat}) \; ^{+0.0005}_{-0.0009}\;(\mathrm{syst})\,.
\end{equation*}

\noindent
From a fit to the \Bs invariant mass distribution in the \mydecay decay, the mass of the \Bs meson is measured to be

\begin{equation*}
 m(B^0_s) = 5367.08 \pm 0.38\; \mathrm{(stat)} \pm 0.15 \; \mathrm{(syst)}\;\mevcc.
\end{equation*}

\noindent
This value is consistent with previous LHCb results~\cite{LHCb-PAPER-2011-035} and with the world average~\cite{PDG2014}.
The overall uncertainty is 20\% larger than the current most precise measurement.
As the systematic uncertainty is a factor of two smaller, further improvement can be expected when larger datasets become available.

\section*{Acknowledgements}

\noindent We express our gratitude to our colleagues in the CERN
accelerator departments for the excellent performance of the LHC. We
thank the technical and administrative staff at the LHCb
institutes. We acknowledge support from CERN and from the national
agencies: CAPES, CNPq, FAPERJ and FINEP (Brazil); NSFC (China);
CNRS/IN2P3 (France); BMBF, DFG and MPG (Germany); INFN (Italy); 
FOM and NWO (The Netherlands); MNiSW and NCN (Poland); MEN/IFA (Romania); 
MinES and FANO (Russia); MinECo (Spain); SNSF and SER (Switzerland); 
NASU (Ukraine); STFC (United Kingdom); NSF (USA).
We acknowledge the computing resources that are provided by CERN, IN2P3 (France), KIT and DESY (Germany), INFN (Italy), SURF (The Netherlands), PIC (Spain), GridPP (United Kingdom), RRCKI and Yandex LLC (Russia), CSCS (Switzerland), IFIN-HH (Romania), CBPF (Brazil), PL-GRID (Poland) and OSC (USA). We are indebted to the communities behind the multiple open 
source software packages on which we depend.
Individual groups or members have received support from AvH Foundation (Germany),
EPLANET, Marie Sk\l{}odowska-Curie Actions and ERC (European Union), 
Conseil G\'{e}n\'{e}ral de Haute-Savoie, Labex ENIGMASS and OCEVU, 
R\'{e}gion Auvergne (France), RFBR and Yandex LLC (Russia), GVA, XuntaGal and GENCAT (Spain), The Royal Society, Royal Commission for the Exhibition of 1851 and the Leverhulme Trust (United Kingdom).

\addcontentsline{toc}{section}{References}

\ifx\mcitethebibliography\mciteundefinedmacro
\PackageError{LHCb.bst}{mciteplus.sty has not been loaded}
{This bibstyle requires the use of the mciteplus package.}\fi
\providecommand{\href}[2]{#2}


\newpage
\input{LHCb_HD_authorlist_2015-06-25}

\end{document}

%% file: preamble.tex
\usepackage{microtype}
\usepackage{lineno}  
\usepackage{xspace} 
\usepackage{caption} 

\usepackage{graphicx}  
\usepackage{color}
\usepackage{colortbl}
\graphicspath{{./figs/}} 

\usepackage{amsmath} 
\usepackage{amssymb}
\usepackage{amsfonts}
\usepackage{upgreek} 

\newcommand*\patchAmsMathEnvironmentForLineno[1]{%
\expandafter\let\csname old#1\expandafter\endcsname\csname #1\endcsname
\expandafter\let\csname oldend#1\expandafter\endcsname\csname
end#1\endcsname
 \renewenvironment{#1}%
   {\linenomath\csname old#1\endcsname}%
   {\csname oldend#1\endcsname\endlinenomath}%
}
\newcommand*\patchBothAmsMathEnvironmentsForLineno[1]{%
  \patchAmsMathEnvironmentForLineno{#1}%
  \patchAmsMathEnvironmentForLineno{#1*}%
}
\AtBeginDocument{%
\patchBothAmsMathEnvironmentsForLineno{equation}%
\patchBothAmsMathEnvironmentsForLineno{align}%
\patchBothAmsMathEnvironmentsForLineno{flalign}%
\patchBothAmsMathEnvironmentsForLineno{alignat}%
\patchBothAmsMathEnvironmentsForLineno{gather}%
\patchBothAmsMathEnvironmentsForLineno{multline}%
\patchBothAmsMathEnvironmentsForLineno{eqnarray}%
}

\usepackage{hyperref}    
\usepackage[all]{hypcap} 

\input{lhcb-symbols-def} 

\usepackage{cite} 
\usepackage{mciteplus}

%% file: lhcb-symbols-def.tex
\def\lhcb {\mbox{LHCb}\xspace}

\def\MagUp {\mbox{\em Mag\kern -0.05em Up}\xspace}

\ifthenelse{\boolean{uprightparticles}}%
{

 \def\Pmu         {\ensuremath{\upmu}\xspace}

 \def\Ppi         {\ensuremath{\uppi}\xspace}

 \def\Pphi        {\ensuremath{\upphi}\xspace}

 \def\Ppsi        {\ensuremath{\uppsi}\xspace}

 \def\PDelta      {\ensuremath{\Delta}\xspace}                 
 \def\PXi      {\ensuremath{\Xi}\xspace}                 
 \def\PLambda      {\ensuremath{\Lambda}\xspace}                 
 \def\PSigma      {\ensuremath{\Sigma}\xspace}                 
 \def\POmega      {\ensuremath{\Omega}\xspace}                 
 \def\PUpsilon      {\ensuremath{\Upsilon}\xspace}                 
 
 \def\PB      {\ensuremath{\mathrm{B}}\xspace}                 
                  
 \def\PD      {\ensuremath{\mathrm{D}}\xspace}

 \def\PJ      {\ensuremath{\mathrm{J}}\xspace}                 
 \def\PK      {\ensuremath{\mathrm{K}}\xspace}

 \def\Pb      {\ensuremath{\mathrm{b}}\xspace}                 
 \def\Pc      {\ensuremath{\mathrm{c}}\xspace}

 \def\Pi      {\ensuremath{\mathrm{i}}\xspace}

 \def\Ps      {\ensuremath{\mathrm{s}}\xspace}

}
{

 \def\Pmu         {\ensuremath{\mu}\xspace}

 \def\Ppi         {\ensuremath{\pi}\xspace}

 \def\Pphi        {\ensuremath{\phi}\xspace}

 \def\Ppsi        {\ensuremath{\psi}\xspace}                 
                  
 \mathchardef\PDelta="7101
 \mathchardef\PXi="7104
 \mathchardef\PLambda="7103
 \mathchardef\PSigma="7106
 \mathchardef\POmega="710A
 \mathchardef\PUpsilon="7107
                  
 \def\PB      {\ensuremath{B}\xspace}                 
                  
 \def\PD      {\ensuremath{D}\xspace}

 \def\PJ      {\ensuremath{J}\xspace}                 
 \def\PK      {\ensuremath{K}\xspace}

 \def\Pb      {\ensuremath{b}\xspace}                 
 \def\Pc      {\ensuremath{c}\xspace}

 \def\Pi      {\ensuremath{i}\xspace}

 \def\Ps      {\ensuremath{s}\xspace}

}

\makeatletter
\ifcase \@ptsize \relax
  \newcommand{\miniscule}{\@setfontsize\miniscule{4}{5}}
\or
  \newcommand{\miniscule}{\@setfontsize\miniscule{5}{6}}
\or
  \newcommand{\miniscule}{\@setfontsize\miniscule{5}{6}}
\fi
\makeatother

\DeclareRobustCommand{\optbar}[1]{\shortstack{{\miniscule (\rule[.5ex]{1.25em}{.18mm})}
  \\ [-.7ex] $#1$}}

\def\muon       {{\ensuremath{\Pmu}}\xspace}

\def\squark    {{\ensuremath{\Ps}}\xspace}

\def\cquark    {{\ensuremath{\Pc}}\xspace}

\def\bquark    {{\ensuremath{\Pb}}\xspace}

\def\pion   {{\ensuremath{\Ppi}}\xspace}

\def\kaon    {{\ensuremath{\PK}}\xspace}
  \def\Kbar    {{\kern 0.2em\overline{\kern -0.2em \PK}{}}\xspace}

\def\KorKbar    {\kern 0.18em\optbar{\kern -0.18em K}{}\xspace}

\def\Kp      {{\ensuremath{\kaon^+}}\xspace}
\def\Km      {{\ensuremath{\kaon^-}}\xspace}

  \def\Dbar    {{\kern 0.2em\overline{\kern -0.2em \PD}{}}\xspace}

\def\DorDbar    {\kern 0.18em\optbar{\kern -0.18em D}{}\xspace}

\def\B       {{\ensuremath{\PB}}\xspace}
\def\Bbar    {{\ensuremath{\kern 0.18em\overline{\kern -0.18em \PB}{}}}\xspace}

\def\BorBbar    {\kern 0.18em\optbar{\kern -0.18em B}{}\xspace}

\def\Bu      {{\ensuremath{\B^+}}\xspace}

\def\Bp      {{\ensuremath{\Bu}}\xspace}

\def\Bs      {{\ensuremath{\B^0_\squark}}\xspace}

\def\jpsi     {{\ensuremath{{\PJ\mskip -3mu/\mskip -2mu\Ppsi\mskip 2mu}}}\xspace}

  \def\Y#1S{\ensuremath{\PUpsilon{(#1S)}}\xspace}

\def\Lbar        {{\ensuremath{\kern 0.1em\overline{\kern -0.1em\PLambda}}}\xspace}
\def\LorLbar    {\kern 0.18em\optbar{\kern -0.18em \PLambda}{}\xspace}

\def\BF         {{\ensuremath{\cal B}}\xspace}

\def\BR         {\BF}
\newcommand{\decay}[2]{\ensuremath{#1\!\to #2}\xspace}         
\def\ra                 {\ensuremath{\rightarrow}\xspace}
\def\to                 {\ensuremath{\rightarrow}\xspace}

\def\BsToJPsiPhi  {\decay{\Bs}{\jpsi\phi}}
\def\BsToJPsiKK   {\decay{\Bs}{\jpsi\kern 0.12em\Kp\Km}}
\def\BsToJPsiPhiPhi {\decay{\Bs}{\jpsi\phi\kern 0.12em\phi}}
\def\BsToJPsiPhiKK  {\decay{\Bs}{\jpsi\phi\kern 0.12em\Kp\Km}}
\def\BsToJPsi4K  {\decay{\Bs}{\jpsi\kern 0.12em\Kp\Km\Kp\Km}}

\def\AT#1     {\ensuremath{A_{\mathrm{T}}^{#1}}\xspace}           

\def\C#1      {\ensuremath{\mathcal{C}_{#1}}\xspace}                       
\def\Cp#1     {\ensuremath{\mathcal{C}_{#1}^{'}}\xspace}                    
\def\Ceff#1   {\ensuremath{\mathcal{C}_{#1}^{\mathrm{(eff)}}}\xspace}        
\def\Cpeff#1  {\ensuremath{\mathcal{C}_{#1}^{'\mathrm{(eff)}}}\xspace}       
\def\Ope#1    {\ensuremath{\mathcal{O}_{#1}}\xspace}                       
\def\Opep#1   {\ensuremath{\mathcal{O}_{#1}^{'}}\xspace}                    

\newcommand{\tev}{\ensuremath{\mathrm{\,Te\kern -0.1em V}}\xspace}
\newcommand{\gev}{\ensuremath{\mathrm{\,Ge\kern -0.1em V}}\xspace}
\newcommand{\mev}{\ensuremath{\mathrm{\,Me\kern -0.1em V}}\xspace}
\newcommand{\kev}{\ensuremath{\mathrm{\,ke\kern -0.1em V}}\xspace}
\newcommand{\ev}{\ensuremath{\mathrm{\,e\kern -0.1em V}}\xspace}
\newcommand{\gevc}{\ensuremath{{\mathrm{\,Ge\kern -0.1em V\!/}c}}\xspace}
\newcommand{\mevc}{\ensuremath{{\mathrm{\,Me\kern -0.1em V\!/}c}}\xspace}
\newcommand{\gevcc}{\ensuremath{{\mathrm{\,Ge\kern -0.1em V\!/}c^2}}\xspace}
\newcommand{\gevgevcccc}{\ensuremath{{\mathrm{\,Ge\kern -0.1em V^2\!/}c^4}}\xspace}
\newcommand{\mevcc}{\ensuremath{{\mathrm{\,Me\kern -0.1em V\!/}c^2}}\xspace}

\def\mum  {\ensuremath{{\,\upmu\rm m}}\xspace}

\def\invfb   {\ensuremath{\mbox{\,fb}^{-1}}\xspace}

\newcommand{\chisqip}{\ensuremath{\chi^2_{\rm IP}}\xspace}

\def\gsim{{~\raise.15em\hbox{$>$}\kern-.85em
          \lower.35em\hbox{$\sim$}~}\xspace}
\def\lsim{{~\raise.15em\hbox{$<$}\kern-.85em
          \lower.35em\hbox{$\sim$}~}\xspace}

\def\sPlot{\mbox{\em sPlot}\xspace}

\def\ptot       {\mbox{$p$}\xspace}
\def\pt         {\mbox{$p_{\rm T}$}\xspace}

\def\dllkpi     {\ensuremath{\mathrm{DLL}_{\kaon\pion}}\xspace}

\def\dllmupi    {\ensuremath{\mathrm{DLL}_{\muon\pi}}\xspace}
\def\dllxpi     {\ensuremath{\mathrm{DLL}_{x\pion}}\xspace}

\def\evtgen     {\mbox{\textsc{EvtGen}}\xspace}

\def\geant      {\mbox{\textsc{Geant4}}\xspace}

\def\photos     {\mbox{\textsc{Photos}}\xspace}

\def\pythia     {\mbox{\textsc{Pythia}}\xspace}

\def\tell1  {TELL1\xspace}
\def\ukl1   {UKL1\xspace}



%% file: title-LHCb-PAPER.tex

\begin{titlepage}
\pagenumbering{roman}

\vspace*{-1.5cm}
\centerline{\large EUROPEAN ORGANIZATION FOR NUCLEAR RESEARCH (CERN)}
\vspace*{1.5cm}
\noindent
\begin{tabular*}{\linewidth}{lc@{\extracolsep{\fill}}r@{\extracolsep{0pt}}}
\ifthenelse{\boolean{pdflatex}}
{\vspace*{-2.7cm}\mbox{\!\!\!\includegraphics[width=.14\textwidth]{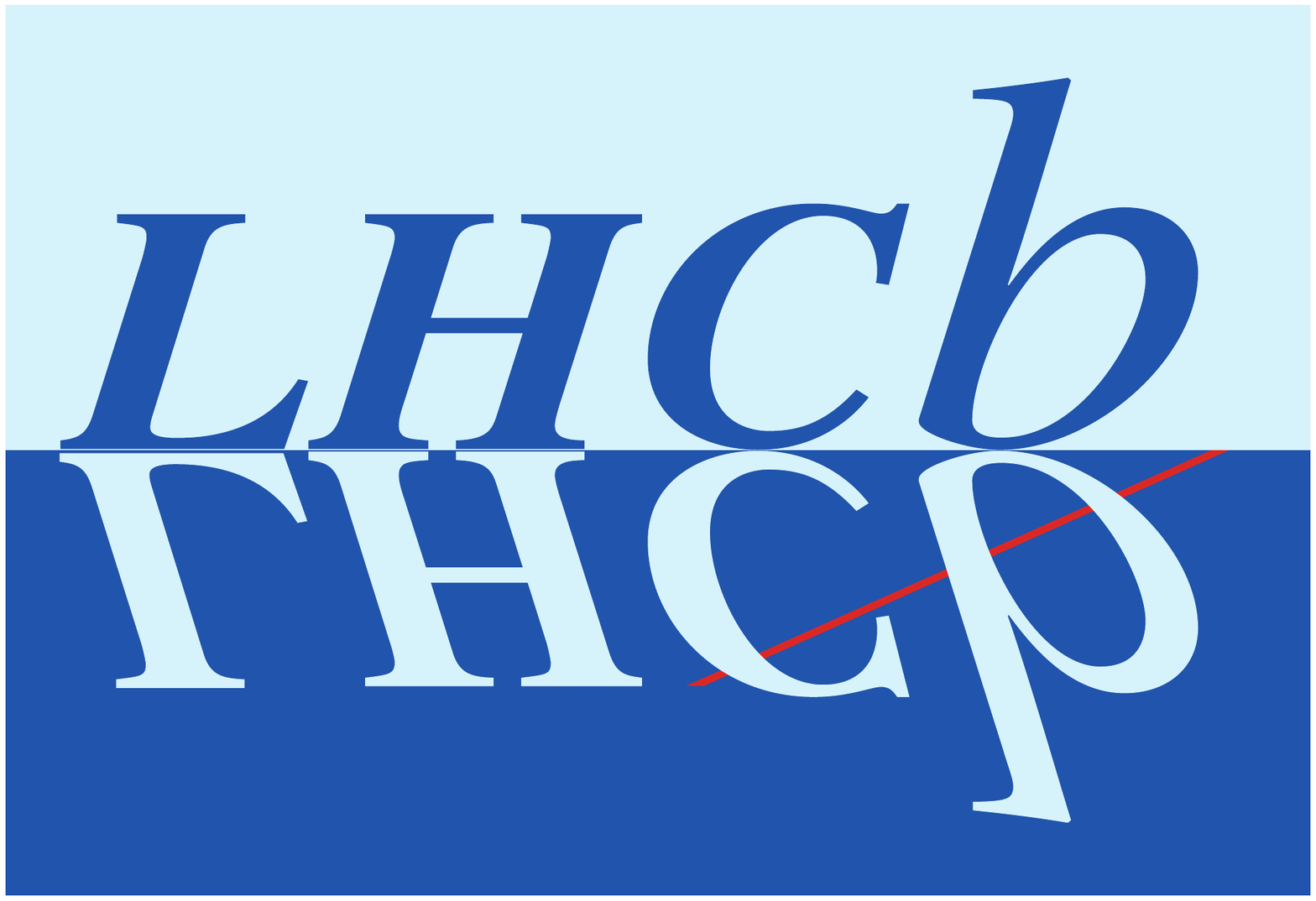}} & &}%
{\vspace*{-1.2cm}\mbox{\!\!\!\includegraphics[width=.12\textwidth]{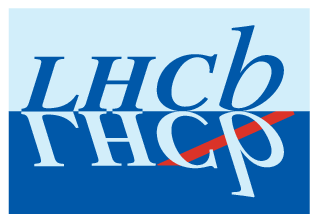}} & &}%
\\
 & & CERN-EP-2016-006 \\  
 & & LHCb-PAPER-2015-033 \\  
 & & January 20, 2016 \\ 
 & & \\
\end{tabular*}

\vspace*{3.0cm}

{\bf\boldmath\LARGE
\begin{center}
  Observation of the \BsToJPsiPhiPhi decay 
\end{center}
}

\vspace*{2.0cm}

\begin{center}
The LHCb collaboration\footnote{Authors are listed at the end of this paper.}
\end{center}

\vspace{\fill}
\begin{abstract}
  \noindent
  The \BsToJPsiPhiPhi decay is observed in $pp$ collision data corresponding to an integrated luminosity of 3\invfb recorded by the LHCb detector at centre-of-mass energies of 7\tev and 8\tev.
  This is the first observation of this decay channel, with a statistical significance of 15 standard deviations.
  The mass of the \Bs meson is measured to be $5367.08\,\! \pm\,\! 0.38\,\! \pm\,\! 0.15\mevcc$.
  The branching fraction ratio $\BR(\BsToJPsiPhiPhi)/\BR(\BsToJPsiPhi)$ is measured to be $0.0115\,\!\pm\,\! 0.0012\; ^{+0.0005}_{-0.0009}$.
  In both cases, the first uncertainty is statistical and the second is systematic. 
  No evidence for non-resonant \BsToJPsiPhiKK  or \BsToJPsi4K decays is found.
\end{abstract}

\vspace*{2.0cm}

\begin{center}
  Published in JHEP {\bf 03} (2016) 040 
\end{center}

\vspace{\fill}

{\footnotesize 
\centerline{\copyright~CERN on behalf of the \lhcb collaboration, licence \href{http://creativecommons.org/licenses/by/4.0/}{CC-BY-4.0}.}}
\vspace*{2mm}

\end{titlepage}


\newpage
\setcounter{page}{2}
\mbox{~}
%
%
%
%

\cleardoublepage

%% file: LHCb_HD_authorlist_2015-06-25.tex
\centerline{\large\bf LHCb collaboration}
\begin{flushleft}
\small
R.~Aaij$^{38}$, 
B.~Adeva$^{37}$, 
M.~Adinolfi$^{46}$, 
A.~Affolder$^{52}$, 
Z.~Ajaltouni$^{5}$, 
S.~Akar$^{6}$, 
J.~Albrecht$^{9}$, 
F.~Alessio$^{38}$, 
M.~Alexander$^{51}$, 
S.~Ali$^{41}$
% ...

{\footnotesize \it
% ...
$ ^{5}$Clermont Universit\'{e}, Universit\'{e} Blaise Pascal, CNRS/IN2P3, LPC, Clermont-Ferrand, France\\
$ ^{6}$CPPM, Aix-Marseille Universit\'{e}, CNRS/IN2P3, Marseille, France\\
% ...
$ ^{9}$Fakult\"{a}t Physik, Technische Universit\"{a}t Dortmund, Dortmund, Germany\\
% ...
$ ^{37}$Universidad de Santiago de Compostela, Santiago de Compostela, Spain\\
$ ^{38}$European Organization for Nuclear Research (CERN), Geneva, Switzerland\\
% ...
$ ^{41}$Nikhef National Institute for Subatomic Physics, Amsterdam, The Netherlands\\
% ...
$ ^{46}$H.H. Wills Physics Laboratory, University of Bristol, Bristol, United Kingdom\\
% ...
$ ^{51}$School of Physics and Astronomy, University of Glasgow, Glasgow, United Kingdom\\
$ ^{52}$Oliver Lodge Laboratory, University of Liverpool, Liverpool, United Kingdom\\
% ...
}
\end{flushleft}

%% file: main.bbl
\begin{mcitethebibliography}{10}
\mciteSetBstSublistMode{n}
\mciteSetBstMaxWidthForm{subitem}{\alph{mcitesubitemcount})}
\mciteSetBstSublistLabelBeginEnd{\mcitemaxwidthsubitemform\space}
{\relax}{\relax}

\bibitem{Lees:2014lra}
BaBar collaboration, J.~P. Lees {\em et~al.},
  \ifthenelse{\boolean{articletitles}}{\emph{{Study of $B^{\pm,0} \to \jpsi K^+
  K^- K^{\pm,0}$ and search for $B^0 \to \jpsi \phi$ at BaBar}},
  }{}\href{http://dx.doi.org/10.1103/PhysRevD.91.012003}{Phys.\ Rev.\
  \textbf{D91} (2015) 012003},
  \href{http://arxiv.org/abs/1407.7244}{{\normalfont\ttfamily
  arXiv:1407.7244}}\relax
\mciteBstWouldAddEndPuncttrue
\mciteSetBstMidEndSepPunct{\mcitedefaultmidpunct}
{\mcitedefaultendpunct}{\mcitedefaultseppunct}\relax
\EndOfBibitem
\bibitem{LHCb-PAPER-2011-033}
LHCb collaboration, R.~Aaij {\em et~al.},
  \ifthenelse{\boolean{articletitles}}{\emph{{Search for the $X(4140)$ state in
  $B^+\to J/\psi\phi K^+$ decays}},
  }{}\href{http://dx.doi.org/10.1103/PhysRevD.85.091103}{Phys.\ Rev.\
  \textbf{D85} (2012) 091103(R)},
  \href{http://arxiv.org/abs/1202.5087}{{\normalfont\ttfamily
  arXiv:1202.5087}}\relax
\mciteBstWouldAddEndPuncttrue
\mciteSetBstMidEndSepPunct{\mcitedefaultmidpunct}
{\mcitedefaultendpunct}{\mcitedefaultseppunct}\relax
\EndOfBibitem
\bibitem{CDF_Y4140}
CDF collaboration, T.~Aaltonen {\em et~al.},
  \ifthenelse{\boolean{articletitles}}{\emph{{Evidence for a narrow
  near-threshold structure in the $\jpsi \phi$ mass spectrum in $B^+ \to \jpsi
  \phi K^+$ decays}},
  }{}\href{http://dx.doi.org/10.1103/PhysRevLett.102.242002}{Phys.\ Rev.\
  Lett.\  \textbf{102} (2009) 242002},
  \href{http://arxiv.org/abs/0903.2229}{{\normalfont\ttfamily
  arXiv:0903.2229}}\relax
\mciteBstWouldAddEndPuncttrue
\mciteSetBstMidEndSepPunct{\mcitedefaultmidpunct}
{\mcitedefaultendpunct}{\mcitedefaultseppunct}\relax
\EndOfBibitem
\bibitem{Gregory:2010gm}
E.~B. Gregory {\em et~al.}, \ifthenelse{\boolean{articletitles}}{\emph{{Precise
  B, $B_s$ and $B_c$ meson spectroscopy from full lattice QCD}},
  }{}\href{http://dx.doi.org/10.1103/PhysRevD.83.014506}{Phys.\ Rev.\
  \textbf{D83} (2011) 014506},
  \href{http://arxiv.org/abs/1010.3848}{{\normalfont\ttfamily
  arXiv:1010.3848}}\relax
\mciteBstWouldAddEndPuncttrue
\mciteSetBstMidEndSepPunct{\mcitedefaultmidpunct}
{\mcitedefaultendpunct}{\mcitedefaultseppunct}\relax
\EndOfBibitem
\bibitem{McNeile:2012qf}
C.~McNeile {\em et~al.}, \ifthenelse{\boolean{articletitles}}{\emph{{Heavy
  meson masses and decay constants from relativistic heavy quarks in full
  lattice QCD}}, }{}\href{http://dx.doi.org/10.1103/PhysRevD.86.074503}{Phys.\
  Rev.\  \textbf{D86} (2012) 074503},
  \href{http://arxiv.org/abs/1207.0994}{{\normalfont\ttfamily
  arXiv:1207.0994}}\relax
\mciteBstWouldAddEndPuncttrue
\mciteSetBstMidEndSepPunct{\mcitedefaultmidpunct}
{\mcitedefaultendpunct}{\mcitedefaultseppunct}\relax
\EndOfBibitem
\bibitem{Lewis:2008fu}
R.~Lewis and R.~M. Woloshyn, \ifthenelse{\boolean{articletitles}}{\emph{{Bottom
  baryons from a dynamical lattice QCD simulation}},
  }{}\href{http://dx.doi.org/10.1103/PhysRevD.79.014502}{Phys.\ Rev.\
  \textbf{D79} (2009) 014502},
  \href{http://arxiv.org/abs/0806.4783}{{\normalfont\ttfamily
  arXiv:0806.4783}}\relax
\mciteBstWouldAddEndPuncttrue
\mciteSetBstMidEndSepPunct{\mcitedefaultmidpunct}
{\mcitedefaultendpunct}{\mcitedefaultseppunct}\relax
\EndOfBibitem
\bibitem{PDG2014}
Particle Data Group, K.~A. Olive {\em et~al.},
  \ifthenelse{\boolean{articletitles}}{\emph{{\href{http://pdg.lbl.gov/}{Review
  of particle physics}}},
  }{}\href{http://dx.doi.org/10.1088/1674-1137/38/9/090001}{Chin.\ Phys.\
  \textbf{C38} (2014) 090001 and 2015 update}\relax
\mciteBstWouldAddEndPuncttrue
\mciteSetBstMidEndSepPunct{\mcitedefaultmidpunct}
{\mcitedefaultendpunct}{\mcitedefaultseppunct}\relax
\EndOfBibitem
\bibitem{LHCb-PAPER-2011-035}
LHCb collaboration, R.~Aaij {\em et~al.},
  \ifthenelse{\boolean{articletitles}}{\emph{{Measurement of $b$-hadron
  masses}}, }{}\href{http://dx.doi.org/10.1016/j.physletb.2012.01.058}{Phys.\
  Lett.\  \textbf{B708} (2012) 241},
  \href{http://arxiv.org/abs/1112.4896}{{\normalfont\ttfamily
  arXiv:1112.4896}}\relax
\mciteBstWouldAddEndPuncttrue
\mciteSetBstMidEndSepPunct{\mcitedefaultmidpunct}
{\mcitedefaultendpunct}{\mcitedefaultseppunct}\relax
\EndOfBibitem
\bibitem{Swanson:2006st}
E.~S. Swanson, \ifthenelse{\boolean{articletitles}}{\emph{{The new heavy
  mesons: A status report}},
  }{}\href{http://dx.doi.org/10.1016/j.physrep.2006.04.003}{Phys.\ Rept.\
  \textbf{429} (2006) 243},
  \href{http://arxiv.org/abs/hep-ph/0601110}{{\normalfont\ttfamily
  arXiv:hep-ph/0601110}}\relax
\mciteBstWouldAddEndPuncttrue
\mciteSetBstMidEndSepPunct{\mcitedefaultmidpunct}
{\mcitedefaultendpunct}{\mcitedefaultseppunct}\relax
\EndOfBibitem
\bibitem{Klempt:2007cp}
E.~Klempt and A.~Zaitsev,
  \ifthenelse{\boolean{articletitles}}{\emph{{Glueballs, hybrids, multiquarks:
  Experimental facts versus QCD inspired concepts}},
  }{}\href{http://dx.doi.org/10.1016/j.physrep.2007.07.006}{Phys.\ Rept.\
  \textbf{454} (2007) 1},
  \href{http://arxiv.org/abs/0708.4016}{{\normalfont\ttfamily
  arXiv:0708.4016}}\relax
\mciteBstWouldAddEndPuncttrue
\mciteSetBstMidEndSepPunct{\mcitedefaultmidpunct}
{\mcitedefaultendpunct}{\mcitedefaultseppunct}\relax
\EndOfBibitem
\bibitem{Shen:2009vs}
Belle collaboration, C.~P. Shen {\em et~al.},
  \ifthenelse{\boolean{articletitles}}{\emph{{Evidence for a new resonance and
  search for the $Y(4140)$ in the $\gamma \gamma \to \phi \jpsi$ process}},
  }{}\href{http://dx.doi.org/10.1103/PhysRevLett.104.112004}{Phys.\ Rev.\
  Lett.\  \textbf{104} (2010) 112004},
  \href{http://arxiv.org/abs/0912.2383}{{\normalfont\ttfamily
  arXiv:0912.2383}}\relax
\mciteBstWouldAddEndPuncttrue
\mciteSetBstMidEndSepPunct{\mcitedefaultmidpunct}
{\mcitedefaultendpunct}{\mcitedefaultseppunct}\relax
\EndOfBibitem
\bibitem{Abazov:2013xda}
D0 collaboration, V.~M. Abazov {\em et~al.},
  \ifthenelse{\boolean{articletitles}}{\emph{{Search for the $X(4140)$ state in
  $B^+\to J/\psi \phi K^+$ decay with the D0 detector}},
  }{}\href{http://dx.doi.org/10.1103/PhysRevD.89.012004}{Phys.\ Rev.\
  \textbf{D89} (2014) 012004},
  \href{http://arxiv.org/abs/1309.6580}{{\normalfont\ttfamily
  arXiv:1309.6580}}\relax
\mciteBstWouldAddEndPuncttrue
\mciteSetBstMidEndSepPunct{\mcitedefaultmidpunct}
{\mcitedefaultendpunct}{\mcitedefaultseppunct}\relax
\EndOfBibitem
\bibitem{Chatrchyan:2013dma}
CMS collaboration, S.~Chatrchyan {\em et~al.},
  \ifthenelse{\boolean{articletitles}}{\emph{{Observation of a peaking
  structure in the $J/\psi \phi$ mass spectrum from $B^{\pm} \to \jpsi \phi
  K^{\pm}$ decays}},
  }{}\href{http://dx.doi.org/10.1016/j.physletb.2014.05.055}{Phys.\ Lett.\
  \textbf{B734} (2014) 261},
  \href{http://arxiv.org/abs/1309.6920}{{\normalfont\ttfamily
  arXiv:1309.6920}}\relax
\mciteBstWouldAddEndPuncttrue
\mciteSetBstMidEndSepPunct{\mcitedefaultmidpunct}
{\mcitedefaultendpunct}{\mcitedefaultseppunct}\relax
\EndOfBibitem
\bibitem{Ablikim2008ac}
BES collaboration, M.~Ablikim {\em et~al.},
  \ifthenelse{\boolean{articletitles}}{\emph{{Partial wave analysis of $\jpsi
  \to \gamma \phi \phi$}},
  }{}\href{http://dx.doi.org/10.1016/j.physletb.2008.03.027}{Phys.\ Lett.\
  \textbf{B662} (2008) 330},
  \href{http://arxiv.org/abs/0801.3885}{{\normalfont\ttfamily
  arXiv:0801.3885}}\relax
\mciteBstWouldAddEndPuncttrue
\mciteSetBstMidEndSepPunct{\mcitedefaultmidpunct}
{\mcitedefaultendpunct}{\mcitedefaultseppunct}\relax
\EndOfBibitem
\bibitem{PhysRevLett.65.1309}
Z.~Bai {\em et~al.}, \ifthenelse{\boolean{articletitles}}{\emph{{Observation of
  a pseudoscalar state in $\jpsi \to \gamma \phi \phi$ near $\phi \phi$
  threshold}}, }{}\href{http://dx.doi.org/10.1103/PhysRevLett.65.1309}{Phys.\
  Rev.\ Lett.\  \textbf{65} (1990) 1309}\relax
\mciteBstWouldAddEndPuncttrue
\mciteSetBstMidEndSepPunct{\mcitedefaultmidpunct}
{\mcitedefaultendpunct}{\mcitedefaultseppunct}\relax
\EndOfBibitem
\bibitem{Bisello1986294}
DM2 collaboration, D.~Bisello {\em et~al.},
  \ifthenelse{\boolean{articletitles}}{\emph{{Search of glueballs in the $\jpsi
  \to \gamma \phi \phi$ decay}},
  }{}\href{http://dx.doi.org/10.1016/0370-2693(86)90584-8}{Phys.\ Lett.\
  \textbf{B179} (1986) 294 }\relax
\mciteBstWouldAddEndPuncttrue
\mciteSetBstMidEndSepPunct{\mcitedefaultmidpunct}
{\mcitedefaultendpunct}{\mcitedefaultseppunct}\relax
\EndOfBibitem
\bibitem{Aubert1}
BaBar collaboration, B.~Aubert {\em et~al.},
  \ifthenelse{\boolean{articletitles}}{\emph{{Rare $B$ decays into states
  containing a \jpsi meson and a meson with $s\bar{s}$ quark content}},
  }{}\href{http://dx.doi.org/10.1103/PhysRevLett.91.071801}{Phys.\ Rev.\ Lett.\
   \textbf{91} (2003) 71801},
  \href{http://arxiv.org/abs/hep-ex/0304014}{{\normalfont\ttfamily
  arXiv:hep-ex/0304014}}\relax
\mciteBstWouldAddEndPuncttrue
\mciteSetBstMidEndSepPunct{\mcitedefaultmidpunct}
{\mcitedefaultendpunct}{\mcitedefaultseppunct}\relax
\EndOfBibitem
\bibitem{Alves:2008zz}
LHCb collaboration, A.~A. Alves~Jr.\ {\em et~al.},
  \ifthenelse{\boolean{articletitles}}{\emph{{The \lhcb detector at the LHC}},
  }{}\href{http://dx.doi.org/10.1088/1748-0221/3/08/S08005}{JINST \textbf{3}
  (2008) S08005}\relax
\mciteBstWouldAddEndPuncttrue
\mciteSetBstMidEndSepPunct{\mcitedefaultmidpunct}
{\mcitedefaultendpunct}{\mcitedefaultseppunct}\relax
\EndOfBibitem
\bibitem{Aaij:2014jba}
LHCb collaboration, R.~Aaij {\em et~al.},
  \ifthenelse{\boolean{articletitles}}{\emph{{\lhcb detector performance}},
  }{}\href{http://dx.doi.org/10.1142/S0217751X15300227}{Int.\ J.\ Mod.\ Phys.\
  \textbf{A30} (2015), no.~07 1530022},
  \href{http://arxiv.org/abs/1412.6352}{{\normalfont\ttfamily
  arXiv:1412.6352}}\relax
\mciteBstWouldAddEndPuncttrue
\mciteSetBstMidEndSepPunct{\mcitedefaultmidpunct}
{\mcitedefaultendpunct}{\mcitedefaultseppunct}\relax
\EndOfBibitem
\bibitem{Sjostrand:2006za}
T.~Sj\"{o}strand, S.~Mrenna, and P.~Skands,
  \ifthenelse{\boolean{articletitles}}{\emph{{PYTHIA 6.4 physics and manual}},
  }{}\href{http://dx.doi.org/10.1088/1126-6708/2006/05/026}{JHEP \textbf{05}
  (2006) 026}, \href{http://arxiv.org/abs/hep-ph/0603175}{{\normalfont\ttfamily
  arXiv:hep-ph/0603175}}\relax
\mciteBstWouldAddEndPuncttrue
\mciteSetBstMidEndSepPunct{\mcitedefaultmidpunct}
{\mcitedefaultendpunct}{\mcitedefaultseppunct}\relax
\EndOfBibitem
\bibitem{LHCb-PROC-2010-056}
I.~Belyaev {\em et~al.}, \ifthenelse{\boolean{articletitles}}{\emph{{Handling
  of the generation of primary events in Gauss, the LHCb simulation
  framework}}, }{}\href{http://dx.doi.org/10.1088/1742-6596/331/3/032047}{{J.\
  Phys.\ Conf.\ Ser.\ } \textbf{331} (2011) 032047}\relax
\mciteBstWouldAddEndPuncttrue
\mciteSetBstMidEndSepPunct{\mcitedefaultmidpunct}
{\mcitedefaultendpunct}{\mcitedefaultseppunct}\relax
\EndOfBibitem
\bibitem{Lange:2001uf}
D.~J. Lange, \ifthenelse{\boolean{articletitles}}{\emph{{The EvtGen particle
  decay simulation package}},
  }{}\href{http://dx.doi.org/10.1016/S0168-9002(01)00089-4}{Nucl.\ Instrum.\
  Meth.\  \textbf{A462} (2001) 152}\relax
\mciteBstWouldAddEndPuncttrue
\mciteSetBstMidEndSepPunct{\mcitedefaultmidpunct}
{\mcitedefaultendpunct}{\mcitedefaultseppunct}\relax
\EndOfBibitem
\bibitem{Golonka:2005pn}
P.~Golonka and Z.~Was, \ifthenelse{\boolean{articletitles}}{\emph{{PHOTOS Monte
  Carlo: A precision tool for QED corrections in $Z$ and $W$ decays}},
  }{}\href{http://dx.doi.org/10.1140/epjc/s2005-02396-4}{Eur.\ Phys.\ J.\
  \textbf{C45} (2006) 97},
  \href{http://arxiv.org/abs/hep-ph/0506026}{{\normalfont\ttfamily
  arXiv:hep-ph/0506026}}\relax
\mciteBstWouldAddEndPuncttrue
\mciteSetBstMidEndSepPunct{\mcitedefaultmidpunct}
{\mcitedefaultendpunct}{\mcitedefaultseppunct}\relax
\EndOfBibitem
\bibitem{Allison:2006ve}
GEANT4 collaboration, J.~Allison {\em et~al.},
  \ifthenelse{\boolean{articletitles}}{\emph{{Geant4 developments and
  applications}}, }{}\href{http://dx.doi.org/10.1109/TNS.2006.869826}{IEEE
  Trans.\ Nucl.\ Sci.\  \textbf{53} (2006) 270}\relax
\mciteBstWouldAddEndPuncttrue
\mciteSetBstMidEndSepPunct{\mcitedefaultmidpunct}
{\mcitedefaultendpunct}{\mcitedefaultseppunct}\relax
\EndOfBibitem
\bibitem{Agostinelli:2002hh}
GEANT4 collaboration, S.~Agostinelli {\em et~al.},
  \ifthenelse{\boolean{articletitles}}{\emph{{GEANT4: A simulation toolkit}},
  }{}\href{http://dx.doi.org/10.1016/S0168-9002(03)01368-8}{Nucl.\ Instrum.\
  Meth.\  \textbf{A506} (2003) 250}\relax
\mciteBstWouldAddEndPuncttrue
\mciteSetBstMidEndSepPunct{\mcitedefaultmidpunct}
{\mcitedefaultendpunct}{\mcitedefaultseppunct}\relax
\EndOfBibitem
\bibitem{LHCb-PROC-2011-006}
M.~Clemencic {\em et~al.}, \ifthenelse{\boolean{articletitles}}{\emph{{The
  \lhcb simulation application, Gauss: Design, evolution and experience}},
  }{}\href{http://dx.doi.org/10.1088/1742-6596/331/3/032023}{{J.\ Phys.\ Conf.\
  Ser.\ } \textbf{331} (2011) 032023}\relax
\mciteBstWouldAddEndPuncttrue
\mciteSetBstMidEndSepPunct{\mcitedefaultmidpunct}
{\mcitedefaultendpunct}{\mcitedefaultseppunct}\relax
\EndOfBibitem
\bibitem{Hulsbergen}
W.~D. Hulsbergen, \ifthenelse{\boolean{articletitles}}{\emph{{Decay chain
  fitting with a Kalman filter}},
  }{}\href{http://dx.doi.org/10.1016/j.nima.2005.06.078}{Nucl.\ Instr.\ Meth.\
  \textbf{A552} (2005) 566},
  \href{http://arxiv.org/abs/physics/0503191}{{\normalfont\ttfamily
  arXiv:physics/0503191}}\relax
\mciteBstWouldAddEndPuncttrue
\mciteSetBstMidEndSepPunct{\mcitedefaultmidpunct}
{\mcitedefaultendpunct}{\mcitedefaultseppunct}\relax
\EndOfBibitem
\bibitem{Breiman}
L.~Breiman, J.~H. Friedman, R.~A. Olshen, and C.~J. Stone, {\em Classification
  and regression trees}, Wadsworth international group, Belmont, California,
  USA, 1984\relax
\mciteBstWouldAddEndPuncttrue
\mciteSetBstMidEndSepPunct{\mcitedefaultmidpunct}
{\mcitedefaultendpunct}{\mcitedefaultseppunct}\relax
\EndOfBibitem
\bibitem{Roe}
B.~P. Roe {\em et~al.}, \ifthenelse{\boolean{articletitles}}{\emph{{Boosted
  decision trees as an alternative to artificial neural networks for particle
  identification}},
  }{}\href{http://dx.doi.org/10.1016/j.nima.2004.12.018}{Nucl.\ Instrum.\
  Meth.\  \textbf{A543} (2005) 577},
  \href{http://arxiv.org/abs/physics/0408124}{{\normalfont\ttfamily
  arXiv:physics/0408124}}\relax
\mciteBstWouldAddEndPuncttrue
\mciteSetBstMidEndSepPunct{\mcitedefaultmidpunct}
{\mcitedefaultendpunct}{\mcitedefaultseppunct}\relax
\EndOfBibitem
\bibitem{AdaBoost}
R.~E. Schapire and Y.~Freund, \ifthenelse{\boolean{articletitles}}{\emph{A
  decision-theoretic generalization of on-line learning and an application to
  boosting}, }{}\href{http://dx.doi.org/10.1006/jcss.1997.1504}{J.\ Comput.\
  Syst.\ Sci.\  \textbf{55} (1997) 119}\relax
\mciteBstWouldAddEndPuncttrue
\mciteSetBstMidEndSepPunct{\mcitedefaultmidpunct}
{\mcitedefaultendpunct}{\mcitedefaultseppunct}\relax
\EndOfBibitem
\bibitem{Punzi:2003bu}
G.~Punzi, \ifthenelse{\boolean{articletitles}}{\emph{{Sensitivity of searches
  for new signals and its optimization}}, }{} in {\em Statistical Problems in
  Particle Physics, Astrophysics, and Cosmology} (L.~{Lyons}, R.~{Mount}, and
  R.~{Reitmeyer}, eds.), p.~79, 2003.
\newblock \href{http://arxiv.org/abs/physics/0308063}{{\normalfont\ttfamily
  arXiv:physics/0308063}}\relax
\mciteBstWouldAddEndPuncttrue
\mciteSetBstMidEndSepPunct{\mcitedefaultmidpunct}
{\mcitedefaultendpunct}{\mcitedefaultseppunct}\relax
\EndOfBibitem
\bibitem{Wilks:1938dza}
S.~S. Wilks, \ifthenelse{\boolean{articletitles}}{\emph{{The large-sample
  distribution of the likelihood ratio for testing composite hypotheses}},
  }{}\href{http://dx.doi.org/10.1214/aoms/1177732360}{Ann.\ Math.\ Stat.\
  \textbf{9} (1938) 60}\relax
\mciteBstWouldAddEndPuncttrue
\mciteSetBstMidEndSepPunct{\mcitedefaultmidpunct}
{\mcitedefaultendpunct}{\mcitedefaultseppunct}\relax
\EndOfBibitem
\bibitem{splot}
M.~Pivk and F.~R. {Le~Diberder},
  \ifthenelse{\boolean{articletitles}}{\emph{{sPlot: A statistical tool to
  unfold data distributions}},
  }{}\href{http://dx.doi.org/10.1016/j.nima.2005.08.106}{Nucl.\ Instrum.\
  Meth.\  \textbf{A555} (2005) 356},
  \href{http://arxiv.org/abs/physics/0402083}{{\normalfont\ttfamily
  arXiv:physics/0402083}}\relax
\mciteBstWouldAddEndPuncttrue
\mciteSetBstMidEndSepPunct{\mcitedefaultmidpunct}
{\mcitedefaultendpunct}{\mcitedefaultseppunct}\relax
\EndOfBibitem
\bibitem{Skwarnicki:1986xj}
T.~Skwarnicki, {\em {A study of the radiative cascade transitions between the
  Upsilon-prime and Upsilon resonances}}, PhD thesis, Institute of Nuclear
  Physics, Krakow, 1986,
  {\href{http://inspirehep.net/record/230779/files/230779.pdf}{DESY-F31-86-02}}\relax
\mciteBstWouldAddEndPuncttrue
\mciteSetBstMidEndSepPunct{\mcitedefaultmidpunct}
{\mcitedefaultendpunct}{\mcitedefaultseppunct}\relax
\EndOfBibitem
\bibitem{TrackingPaper}
LHCb collaboration, R.~Aaij {\em et~al.},
  \ifthenelse{\boolean{articletitles}}{\emph{{Measurement of the track
  reconstruction efficiency at \lhcb}},
  }{}\href{http://dx.doi.org/10.1088/1748-0221/10/02/P02007}{JINST \textbf{10}
  (2015) P02007}, \href{http://arxiv.org/abs/1408.1251}{{\normalfont\ttfamily
  arXiv:1408.1251}}\relax
\mciteBstWouldAddEndPuncttrue
\mciteSetBstMidEndSepPunct{\mcitedefaultmidpunct}
{\mcitedefaultendpunct}{\mcitedefaultseppunct}\relax
\EndOfBibitem
\bibitem{FeldmanCousins}
G.~J. Feldman and R.~D. Cousins,
  \ifthenelse{\boolean{articletitles}}{\emph{{Unified approach to the classical
  statistical analysis of small signals}},
  }{}\href{http://dx.doi.org/10.1103/PhysRevD.57.3873}{Phys.\ Rev.\
  \textbf{D57} (1998) 3873},
  \href{http://arxiv.org/abs/physics/9711021}{{\normalfont\ttfamily
  arXiv:physics/9711021}}\relax
\mciteBstWouldAddEndPuncttrue
\mciteSetBstMidEndSepPunct{\mcitedefaultmidpunct}
{\mcitedefaultendpunct}{\mcitedefaultseppunct}\relax
\EndOfBibitem
\bibitem{DeBruyn:2012wj}
K.~De~Bruyn {\em et~al.}, \ifthenelse{\boolean{articletitles}}{\emph{{Branching
  ratio measurements of $B_s$ decays}},
  }{}\href{http://dx.doi.org/10.1103/PhysRevD.86.014027}{Phys.\ Rev.\
  \textbf{D86} (2012) 014027},
  \href{http://arxiv.org/abs/1204.1735}{{\normalfont\ttfamily
  arXiv:1204.1735}}\relax
\mciteBstWouldAddEndPuncttrue
\mciteSetBstMidEndSepPunct{\mcitedefaultmidpunct}
{\mcitedefaultendpunct}{\mcitedefaultseppunct}\relax
\EndOfBibitem
\bibitem{Bs_JpsiPhi}
LHCb collaboration, R.~Aaij {\em et~al.},
  \ifthenelse{\boolean{articletitles}}{\emph{{Precision measurement of CP
  violation in $B^0_s \to \jpsi K^+ K^-$ decays}},
  }{}\href{http://dx.doi.org/10.1103/PhysRevLett.114.041801}{Phys.\ Rev.\
  Lett.\  \textbf{114} (2015) 041801},
  \href{http://arxiv.org/abs/1411.3104}{{\normalfont\ttfamily
  arXiv:1411.3104}}\relax
\mciteBstWouldAddEndPuncttrue
\mciteSetBstMidEndSepPunct{\mcitedefaultmidpunct}
{\mcitedefaultendpunct}{\mcitedefaultseppunct}\relax
\EndOfBibitem
\bibitem{TrackingScale}
LHCb collaboration, R.~Aaij {\em et~al.},
  \ifthenelse{\boolean{articletitles}}{\emph{{Precision measurement of D meson
  mass differences}}, }{}\href{http://dx.doi.org/10.1007/JHEP06(2013)065}{JHEP
  \textbf{06} (2013) 065},
  \href{http://arxiv.org/abs/1304.6865}{{\normalfont\ttfamily
  arXiv:1304.6865}}\relax
\mciteBstWouldAddEndPuncttrue
\mciteSetBstMidEndSepPunct{\mcitedefaultmidpunct}
{\mcitedefaultendpunct}{\mcitedefaultseppunct}\relax
\EndOfBibitem
\end{mcitethebibliography}
